# Conoscopic Patterns for Optically Uniaxial Gyrotropic Crystals in the Vicinity of Isotropic Point

Yu. Vasylkiv, Yu. Nastishin and R. Vlokh

Institute of Physical Optics, 23 Dragomanov St., 79005 Lviv, Ukraine,
e-mail: vlokh@ifo.lviv.ua

## Abstract

We have presented computer simulations of conoscopic patterns occurring in optically uniaxial gyrotropic crystals in the vicinity of isotropic point for a number of sets of crystal parameters and orientations. The appearance of special directions characterized by equalization of linear and circular birefringences has been revealed.

## Introduction

The term "isotropic point" concerns the conditions, under which optical birefringence of a non-cubic, optically uniaxial crystal becomes zero. Such the conditions are usually realized at a certain wavelength $\lambda_i$ of optical radiation and they imply in fact a reversal of the linear birefringence sign. One of the features of the isotropic point is that the crystal remains anisotropic at $\lambda_i$, thus offering a unique possibility for observation of anisotropy of the other optical effects, for instance, the optical activity (see, e.g., [1–5]). It is well known that the indicative surface of the optical activity describing the second-rank axial tensor $g_{ij}$ can be presented as a pseudo scalar gyration parameter:

$$G = g_{ij} l_i l_j, \qquad (1)$$

where $l_1 = \sin\Theta\cos\varphi$, $l_2 = \sin\Theta\sin\varphi$ and $l_3 = \cos\Theta$ define transformation relations between the polar and Cartesian coordinate systems and the unit vector $\vec{r}$ coincides with the wave vector (see

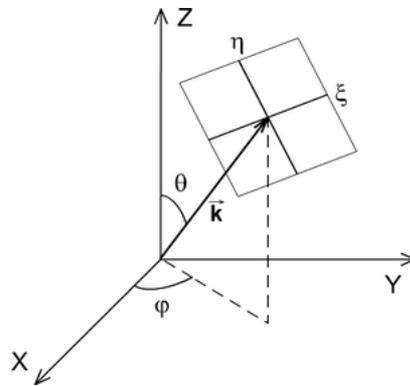

**Fig. 1.** Relation between the Cartesian and polar coordinate systems and orientation of crystal plate with respect to the optical ray.

Fig. 1) [6]. According to one of the properties of the indicative surfaces of $g_{ij}$ tensor, these surfaces correspond to the surfaces of circular birefringence:

$$\Delta n_c = n_r - n_l = \frac{\rho\lambda}{\pi} = \frac{G}{\bar{n}}, \qquad (2)$$

where $n_r$ and $n_l$ are the refractive indices of right and left circularly polarized waves, $\rho$ the specific rotatory power and $\bar{n}$ the mean refractive index of linearly polarized waves peculiar for the given direction of light propagation. The birefringence in gyrotropic crystals with the isotropic point does not completely disappear at $\lambda_i$, because the circular birefringence remains to exist. Moreover, the circular birefringence should be characterized by some spatial distribution, since the crystal remains anisotropic at $\lambda_i$. In particular, the distribution of the circular birefringence should lead to appearance of conoscopic fringes [7-9]. As we have earlier shown [7–9], the conoscopic fringes caused by spatial distribution of the circular birefringence differ from those caused by the linear birefringence. The observation of "circular conoscopic fringes" represents rather difficult experimental task because the circular birefringence effect is usually several orders of magnitude smaller than the linear one. Really, the usual value of the circular birefringence is $\sim 10^{-5}$, while the same for its linear analogue is of the order of $10^{-3}$. It means that, in order to achieve the same spatial density of the interference fringes as in case of the linear birefringence, it would be necessary to provide a sample with the linear size along the light propagation direction, which is two orders of magnitude larger than for the common case. However, if we approach the isotropic point while varying the light wavelength, we may suppose that there would be a spectral range in the vicinity of this point, where the circular and linear birefringences are of the same order. Then the optical retardation should be determined by a total birefringence:

$$\Delta n = (\Delta n_l^2 + \Delta n_c^2)^{1/2}, \qquad (3)$$

where $\Delta n_l$ is the linear birefringence. It would be natural to assume that the conoscopic fringes in this spectral range are determined by spatial distribution of the total birefringence (or the birefringence for elliptically polarized eigenwaves) and, thus, the shape of the fringes should differ from those caused by the pure (or quasi-pure) linear or circular birefringences.

The aim of the present paper is to analyze the shape of conoscopic fringes for optically uniaxial gyrotropic crystals in the close spectral range of the isotropic point.

**Simulation Algorithm**

To help in understanding of the results presented below, let us recall in brief the known algorithm [10] for the calculations of conoscopic patterns in the case of birefringent gyrotropic crystals. The optical transmittance (the light intensity normalized to its value at the entrance interface) of a non-absorbing gyrotropic birefringent crystal plate placed between crossed polarizers takes the following form [11]:

$$I = \cos^2\chi - \sin 2\beta \sin 2(\beta - \chi)\sin^2\frac{\delta}{2}, \qquad (4)$$

where

$$\delta = \frac{2\pi}{\lambda}\Delta nd \qquad (5)$$

is the phase difference (with $\Delta n$ being determined by Eq. (3)), $\chi$ represents the angle defining the off-orientation of polarizer and analyzer and $\beta$ the angle between one of crystallophysical directions lying in the plane perpendicular to the optical beam and the transmittance direction of polarizer. For our geometry ($\chi = \pi/2$ and $\beta = \pi/4$), we can rewrite Eq. (4) as follows:

$$I = \sin^2 \frac{\pi}{\lambda} d \sqrt{\Delta n_l^2 + \frac{G^2}{n^2}}.\qquad(6)$$

Let us remind that Eq. (6) is written only for the central beam of the divergent ray. Accounting for spatial distribution of the linear and circular birefringences modifies Eq. (6), resulting in

$$I = \sin^2 \frac{\pi}{\lambda} d \sqrt{\Delta n_m^2 \sin^2 \Theta + \frac{g_{ij}^2 l_i^2 l_j^2}{n^2}},\qquad(7)$$

where $\Delta n_m$ denotes the maximum value of the linear birefringence in uniaxial crystals for the propagation direction perpendicular to the optic axis. Finally, after considering the angular dependence of $d$, we obtain the relation

$$I = \sin^2 \frac{\pi}{\lambda} \times$$
$$\times \sqrt{(d^2 + \eta^2 + \xi^2)\left(\Delta n_m^2 \sin^2 \Theta + \frac{g_{ij}^2 l_i^2 l_j^2}{n^2}\right)},\qquad(8)$$

where $\eta$ and $\xi$ are the coordinates of the exit interface of the crystal plate (the origin of the coordinate system $\eta, \xi$ coincides with the output point for the central beam of the ray cone – see Fig. 1). Prior to description of the obtained results, it would be reasonable to present a few definitions used later:

(i) the optic axis for linearly polarized waves is a direction in crystal, in which the velocities (or the corresponding refractive indices) of orthogonal linearly polarized eigenwaves are the same;

(ii) the optic axis for circularly polarized waves is a direction in anisotropic crystal at the wavelength of the isotropic point, in which the velocities (or the refractive indices) of left and right circularly polarized eigenwaves are the same (we have $G = 0$ for these directions).

**Results and discussion**

*a) Propagation of light along z axis in crystals belonging to the point symmetry groups 622, 6, 32, 3, 422 and 4.*

The results of computer simulations of the conoscopic patterns for the case of propagation of light along z direction in crystals that belong to the point groups of symmetry 622, 6, 32, 3, 422 and 4 are presented in Figures 2 to 4. The gyration tensor for these symmetry groups takes the form

$$g_{ij} = \begin{vmatrix} g_{11} & 0 & 0 \\ 0 & g_{11} & 0 \\ 0 & 0 & g_{33} \end{vmatrix} \text{ or}$$

$$g_{ij} = \begin{vmatrix} -g_{11} & 0 & 0 \\ 0 & -g_{11} & 0 \\ 0 & 0 & g_{33} \end{vmatrix}.\qquad(9)$$

Hence, the pseudo scalar gyration parameter may be represented as

$$G = g_{11}\sin^2\Theta\cos^2\varphi + g_{11}\sin^2\Theta\sin^2\varphi + g_{33}\cos^2\Theta = g_{11}\sin^2\Theta + g_{33}\cos^2\Theta,$$
(10)

or

$$G = -g_{11}\sin^2\Theta\cos^2\varphi - g_{11}\sin^2\Theta\sin^2\varphi + g_{33}\cos^2\Theta = g_{33}\cos^2\Theta - g_{11}\sin^2\Theta.$$
(11)

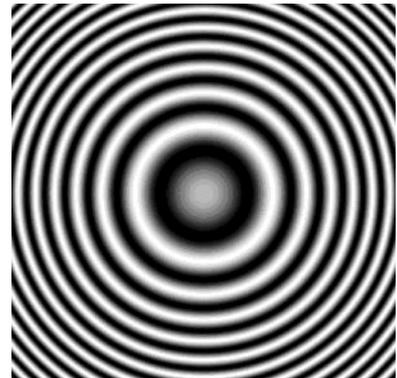

**Fig. 2.** Conoscopic patterns for gyrotropic crystals in case of $g_{11} = g_{22} = g_{33} \neq 0$ and $\Delta n_l = 0$ ($\lambda = 632.8 nm$ and $d = 1m$).

We define the aperture angle of the divergent optical ray to be equal to $45°$. Let the crystal plate have the thickness $d = 1m$ (such artificially large value is used for increasing the circular retardation) and the gyration tensor components are $5\times 10^{-5}$. We will change the linear birefringence in the range of $0-10^{-4}$. As a first step, we demonstrate the conoscopic fringes for the simplest case, when $g_{11} = g_{22} = g_{33} \neq 0$ and $\Delta n_l = 0$ (see Fig. 2). Here the appearance of interference patterns is caused merely by angular dependence of the retardation on the sample thickness. When $g_{11} = g_{22} = g_{33} = 0$ and $\Delta n_l = 0$, the dark field would be observed on the screen.

In the next two cases the gyration tensor surfaces do not exhibit a sign reversal and so there is no direction in crystal, where the velocities of the right and left circularly polarized waves are equal. Increase in the linear birefringence in those cases (see Fig. 3 and 4) leads only to increasing density of ring-like conoscopic fringes.

Gradual changes in the conoscopic fringes for the crystals with sign-reversed gyration surface occurring under increasing linear birefringence exhibit peculiarities consisting in aperiodicity, i.e. the appearance of thick interference ring that moves towards the centre of the figure (see Fig. 5 and 6). The ring appears due to equalization of the linear and circular birefringence values. Moreover, one can see from Fig. 5 and 6 that the surfaces of total elliptical birefringence reveal special directions characterized by extreme values of the latter parameter. The angular dependence of the total birefringence in the vicinity of those directions is not monotonous and the same holds true of the corresponding conoscopic fringes. Furthermore, the circular and linear birefringence values in those directions are the same.

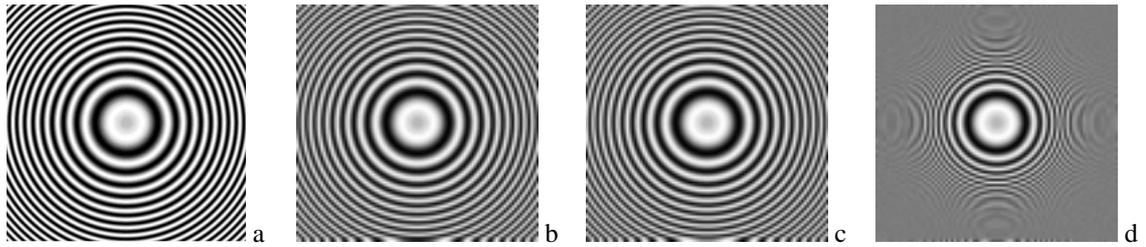

**Fig. 3.** Conoscopic patterns for gyrotropic crystals at the light propagation along z- axis at different values of linear birefringence: a) $\Delta n_l = 0$, b) $\Delta n_l = 10^{-6}$, c) $\Delta n_l = 10^{-5}$, d) $\Delta n_l = 10^{-4}$ ($g_{11} = g_{22} = 5 \times 10^{-5}$, $g_{33} = 2.5 \times 10^{-5}$, $\lambda = 632.8 nm$, $d = 1m$).

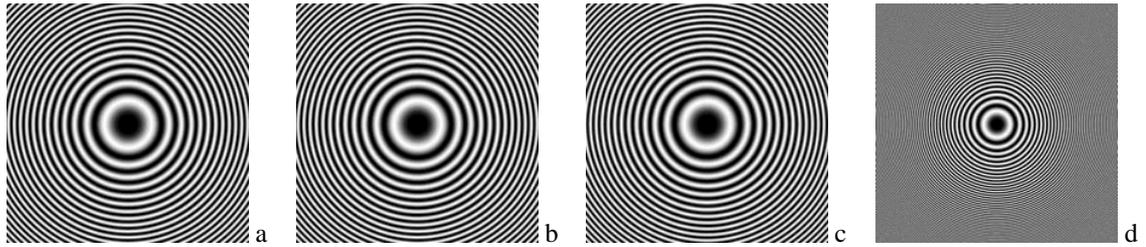

**Fig. 4.** Conoscopic patterns of gyrotropic crystals at the light propagation along z-axis at different values of linear birefringence: a) $\Delta n_l = 0$, b) $\Delta n_l = 10^{-6}$, c) $\Delta n_l = 10^{-5}$, d) $\Delta n_l = 10^{-4}$ ($g_{11} = 10^{-5}$, $g_{33} = 0$, $\lambda = 632.8 nm$, $d = 1m$).

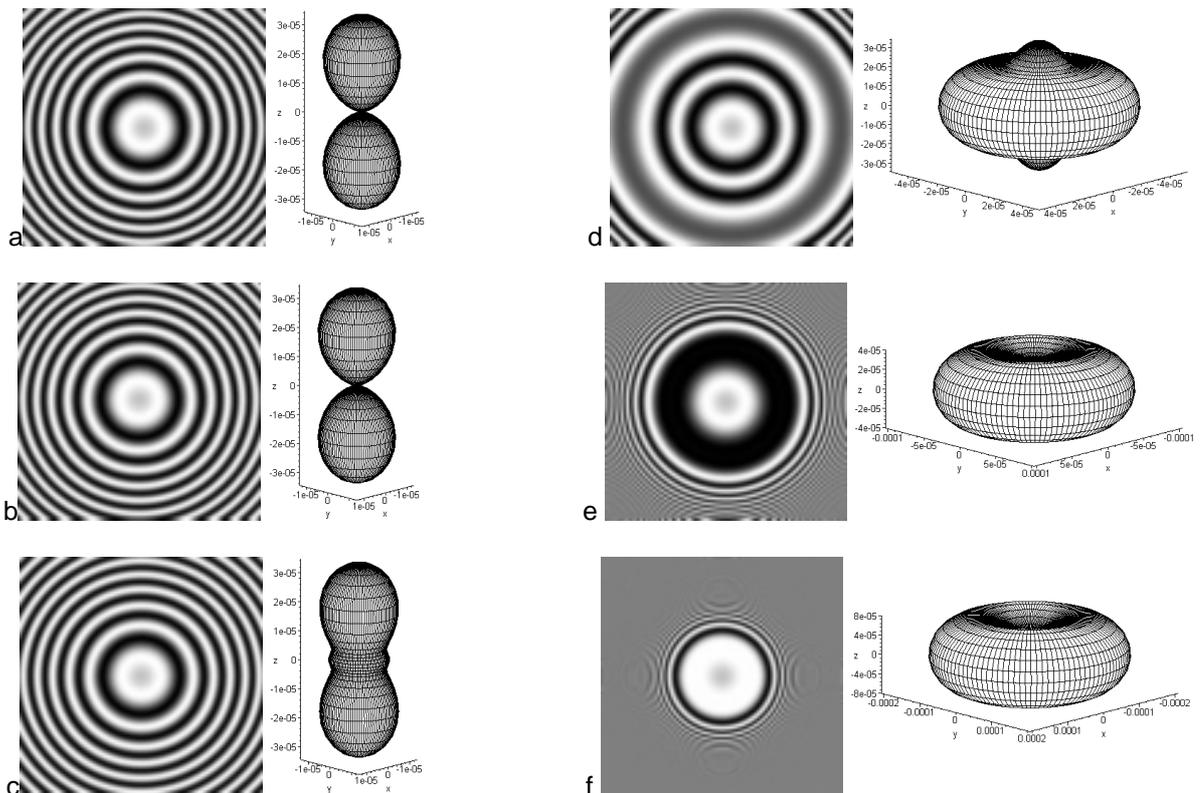

**Fig. 5.** Conoscopic patterns of gyrotropic crystals at the light propagation along z- axis at different values of linear birefringence and surfaces of total elliptical birefringence: a) $\Delta n_l = 0$, b) $\Delta n_l = 10^{-6}$, c) $\Delta n_l = 10^{-5}$, d) $\Delta n_l = 5 \times 10^{-5}$, e) $\Delta n_l = 10^{-4}$, f) $\Delta n_l = 2 \times 10^{-4}$ ($g_{11} = 0$, $g_{33} = 5 \times 10^{-5}$, $\lambda = 632.8 nm$, $d = 1m$).

*b) Propagation of light perpendicular to z axis in crystals belonging to the point symmetry groups 622, 6, 32, 3, 422 and 4.*

In case of the gyration surface showing no sign reversal, the conoscopic fringes are the same as the usual conoscopic fringes caused only by the linear birefringence in optically uniaxial crystals and observed in the direction of light propagation perpendicular to the optic axis (see Fig. 7).

However, the changes in the conoscopic patterns are essentially different in the other cases. For example, a wide interference band is observed under the conditions $g_{11}=g_{22}=0$, $g_{33}=5\times10^{-5}$ and $\Delta n_l=0$ (Fig. 8a). This band is perpendicular to z axis. It is noteworthy that the xy plane is then the plane of absence of circular birefringence or, in some other terms, the plane of optic axis for the circularly polarized waves. With increasing linear birefringence, this band is transformed to oval (see Fig. 8c and d), which is peculiar for the light propagation along the bisector of acute angle between the optic axes in optically biaxial crystals. The angle between those "optic axes" increases with subsequent increase in the linear birefringence (Fig. 8e). The directions of the "optic axes" form the cone, whose angle decreases with increasing linear birefringence. The section of this cone by the xy plane represents a thick interference ring seen in Fig. 5d and e. Finally, the cone angle (here the matter concerns the cone of special directions characterized by extreme values of the total elliptical birefringence and equal values of the linear and circular birefringences) tends to the critical angle, whose magnitude is determined by the ratio of linear and circular birefringences (Fig. 5f).

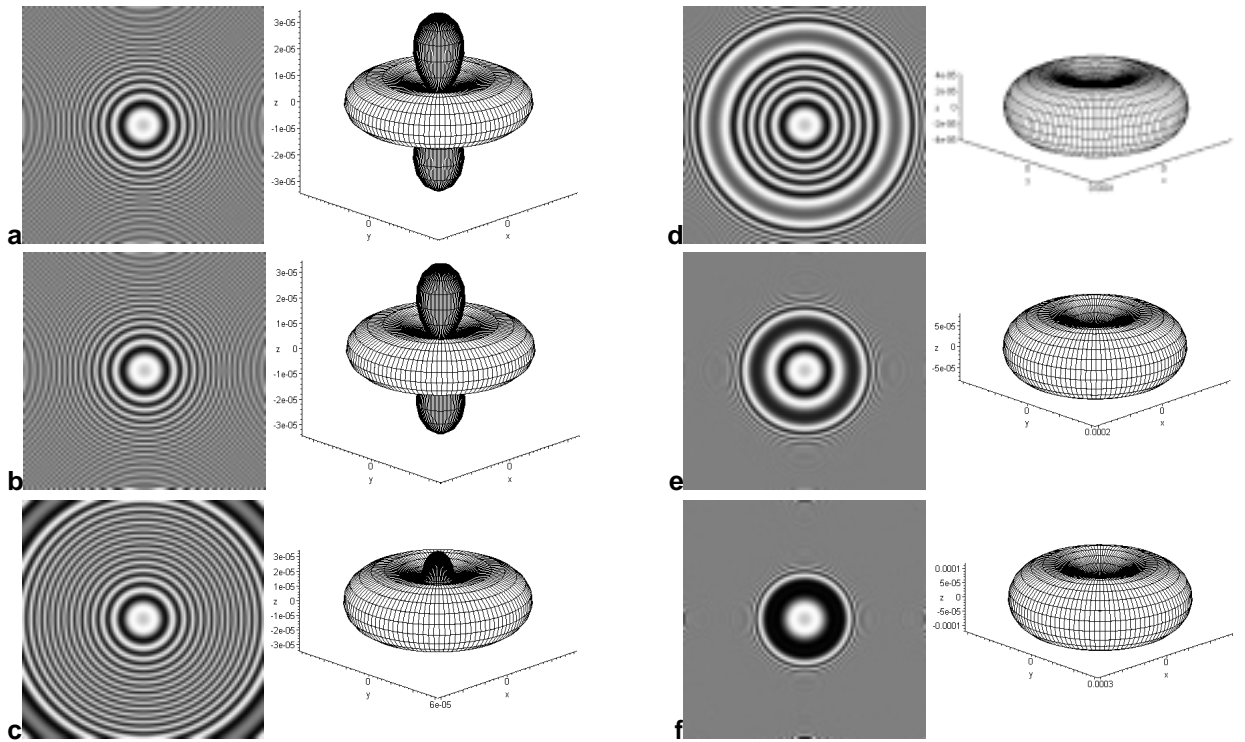

**Fig. 6.** Conoscopic patterns for the propagation direction along z axis in gyrotropic crystals at different values of linear birefringence and the corresponding surfaces of total elliptical birefringence: (a) $\Delta n_l = 0$, (b) $\Delta n_l = 10^{-5}$, (c) $\Delta n_l = 5\times10^{-5}$, (d) $\Delta n_l = 10^{-4}$, (e) $\Delta n_l = 2\times10^{-4}$ and (f) $\Delta n_l = 3\times10^{-4}$ ($g_{11} = 5\times10^{-5}$, $g_{33} = -5\times10^{-5}$, $\lambda = 632.8nm$ and $d = 1m$).

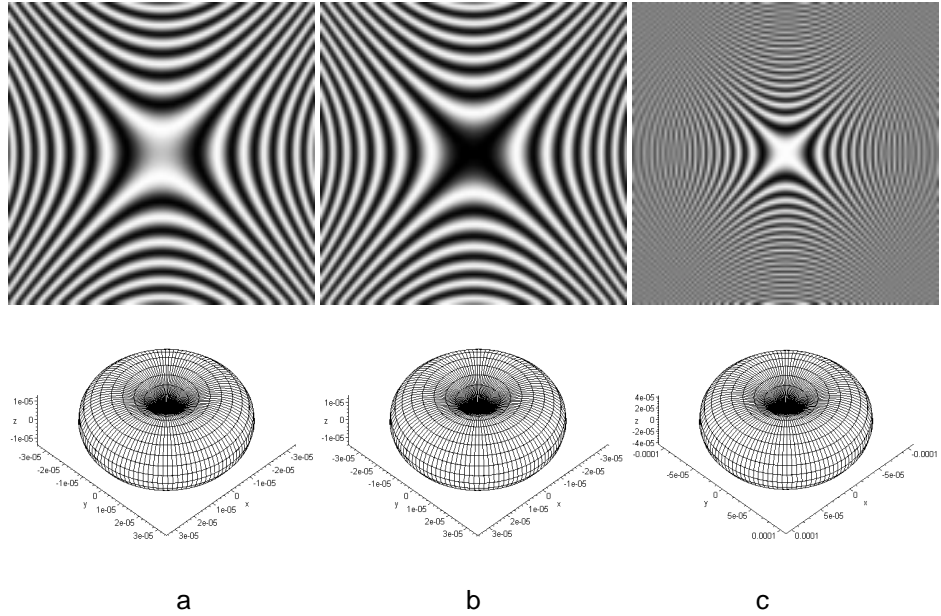

**Fig. 7.** Conoscopic patterns for the propagation direction perpendicular to the optic axis in gyrotropic crystals at different values of linear birefringence: (a) $\Delta n_l = 0$, (b) $\Delta n_l = 10^{-5}$ and (c) $\Delta n_l = 10^{-4}$ ($g_{11} = 5 \times 10^{-5}$, $g_{33} = 0$, $\lambda = 632.8 nm$ and $d = 1m$).

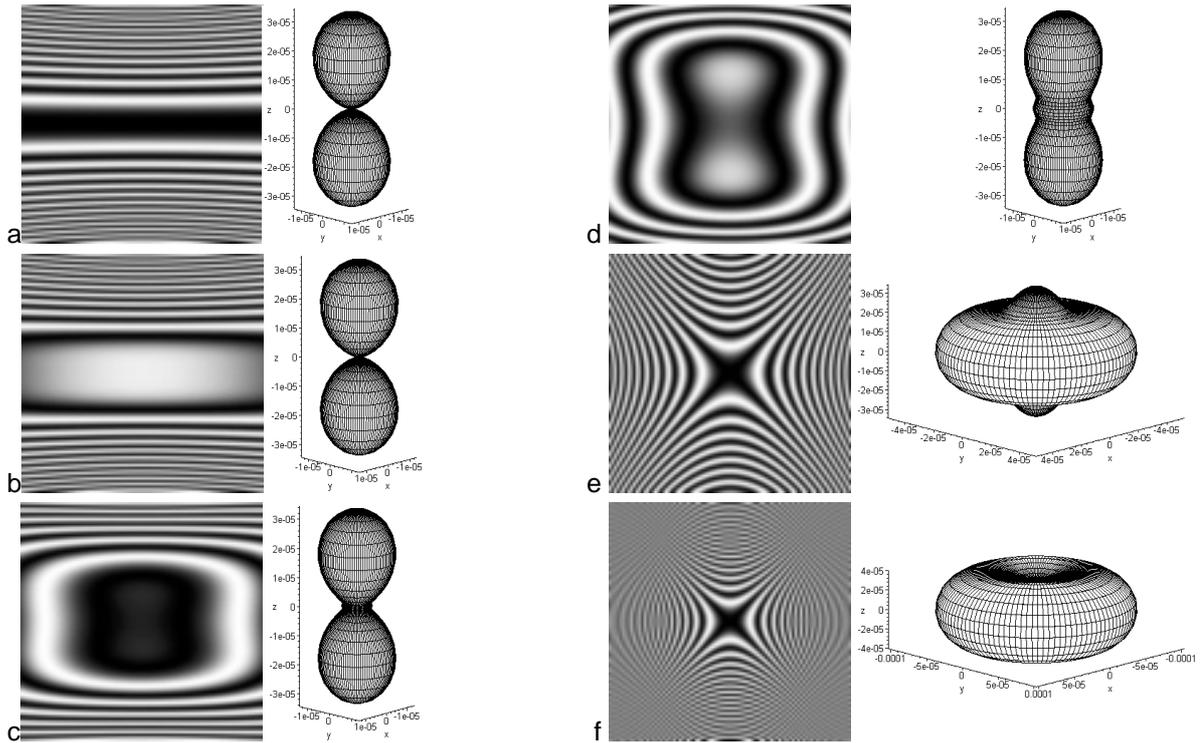

**Fig. 8.** Conoscopic patterns for the propagation direction perpendicular to z axis in gyrotropic crystals at different values of linear birefringence and the corresponding surfaces of total elliptical birefringence: (a) $\Delta n_l = 0$, (b) $\Delta n_l = 10^{-6}$, (c) $\Delta n_l = 5 \times 10^{-6}$, (d) $\Delta n_l = 10^{-5}$, (e) $\Delta n_l = 5 \times 10^{-5}$ and (f) $\Delta n_l = 10^{-4}$ ($g_{11} = 0$, $g_{33} = 5 \times 10^{-5}$, $\lambda = 632.8 nm$ and $d = 1m$).

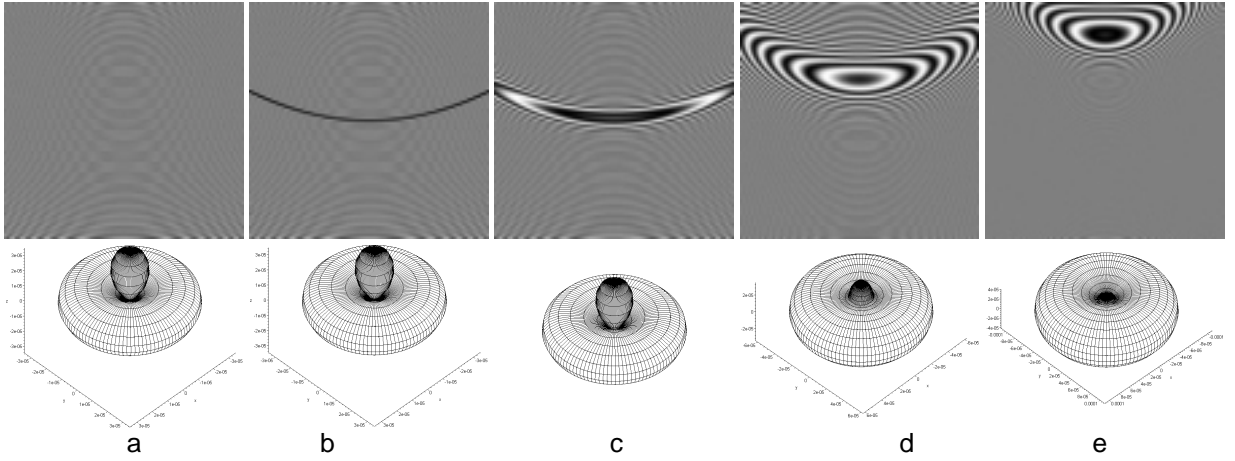

**Fig. 9.** Conoscopic patterns for the propagation direction characterized with the conditions $G = \Delta n_c = 0$ ($45°$ with respect of z axis in the zx plane) and perpendicular to the optic axis in gyrotropic crystals at different values of linear birefringence and the corresponding surfaces of total elliptical birefringence: (a) $\Delta n_l = 0$, (b) $\Delta n_l = 10^{-6}$, (c) $\Delta n_l = 10^{-5}$, (d) $\Delta n_l = 5 \times 10^{-5}$ and (e) $\Delta n_l = 10^{-4}$ ($g_{11} = 5 \times 10^{-5}$, $g_{33} = -5 \times 10^{-5}$, $\lambda = 632.8 nm$ and $d = 1m$).

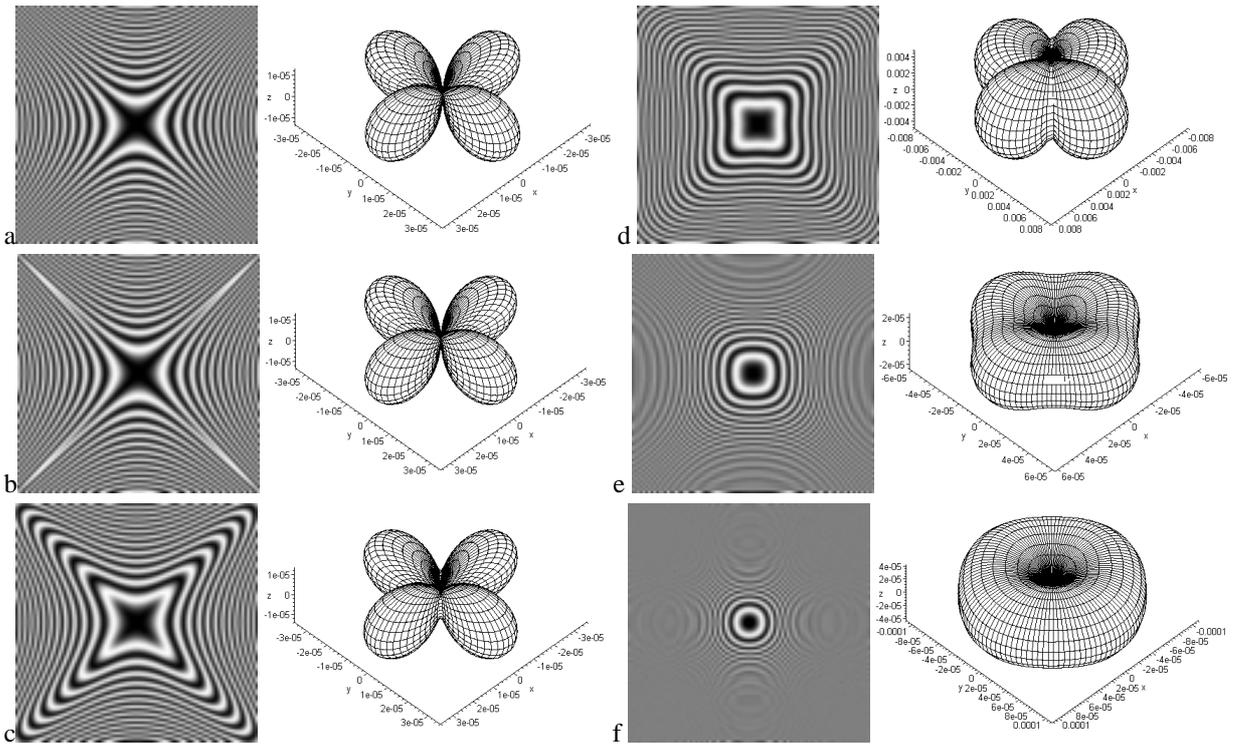

**Fig. 10.** Conoscopic patterns for the propagation direction along z direction in gyrotropic crystals of point group $\bar{4}2m$ at different values of linear birefringence and the corresponding surfaces of total elliptical birefringence: (a) $\Delta n_l = 0$, (b) $\Delta n_l = 10^{-6}$, (c) $\Delta n_l = 10^{-5}$, (d) $\Delta n_l = 2.5 \times 10^{-5}$, (e) $\Delta n_l = 5 \times 10^{-5}$ and (f) $\Delta n_l = 10^{-4}$ ($g_{11} = 5 \times 10^{-5}$, $\lambda = 632.8 nm$ and $d = 1m$).

Here we are to remind that the eigenwaves dealt with are elliptically polarized, whenever the circular and linear birefringences simultaneously exist. Thus, one should consider the refractive indices and the birefringence for those elliptically polarized waves. The changes in the conoscopic patterns presented in Fig. 8a–d correspond to the conditions, under which the direction of equality of the refractive indices for the right and left circularly polarized waves (it is in the xy plane – see Fig. 8a) transforms gradually to the cone of special directions, which are characterized by anomalous values of the total elliptical birefringence (see Fig. 8c and d). Further increase in the linear birefringence gives rise to decreasing cone angle. Similar behaviour is observed in case when the light propagates along the direction of zero gyration parameter (i.e., a zero circular birefringence) in crystals with the sign-reversed gyration surface.

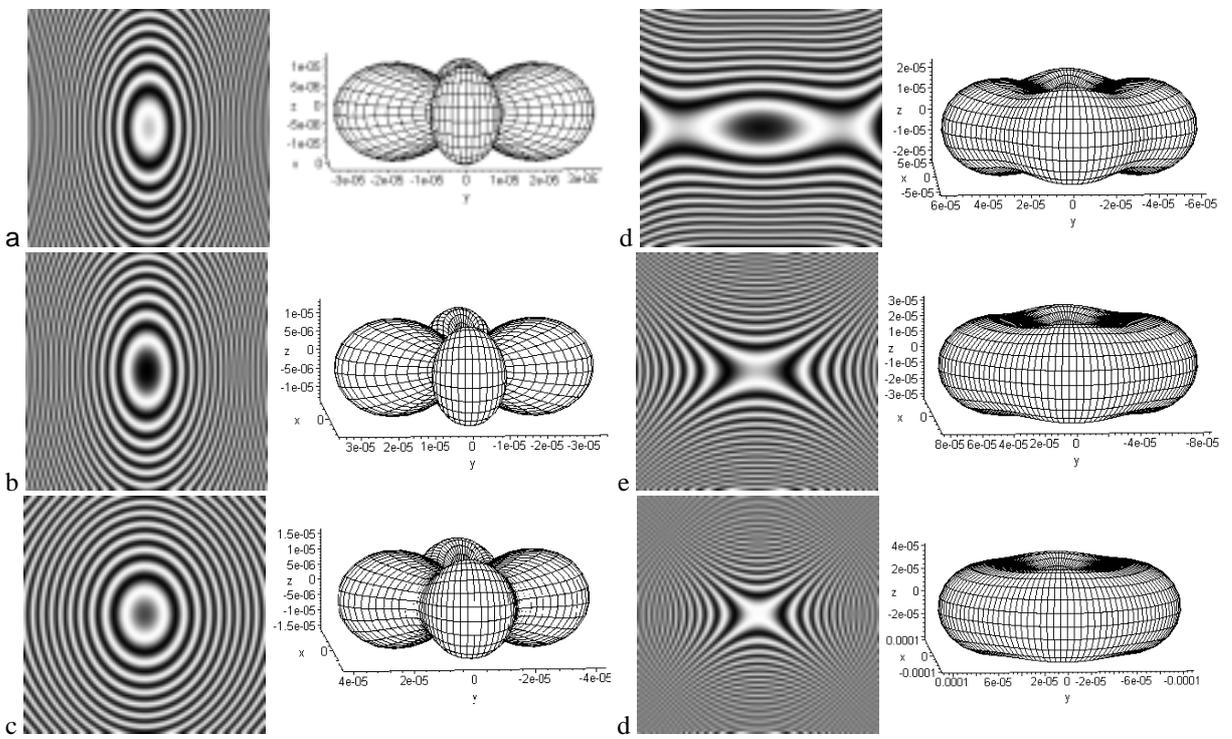

**Fig. 11.** Conoscopic patterns for the propagation direction along x or y directions in gyrotropic crystals of point group $\bar{4}2m$ at different values of linear birefringence and the corresponding surfaces of total elliptical birefringence: (a) $\Delta n_l = 0$, (b) $\Delta n_l = 10^{-5}$, (c) $\Delta n_l = 2.5 \times 10^{-5}$, (d) $\Delta n_l = 5 \times 10^{-5}$, (e) $\Delta n_l = 7.5 \times 10^{-5}$ and (f) $\Delta n_l = 10^{-4}$ ( $g_{11} = 5 \times 10^{-5}$, $\lambda = 632.8nm$ and $d = 1m$).

*c) Propagation of light in the direction where $G = \Delta n_c = 0$ in the uniaxial crystals of point symmetry groups 622, 6, 32, 3, 422 and 4.*

In Fig. 9 one can see how the special directions appear when the light propagates in zx plane along the $45°$-direction with respect to z axis. With increasing linear birefringence, the angle of the cone, where these special directions belong, nears to the smallest value determined by the ratio of the circular and linear birefringences (see e.g [12]) (Fig. 6c–f)

*d) Propagation of light along the optic axis in the uniaxial crystals of point symmetry group $\bar{4}2m$.*

The gyration tensor for the point group of symmetry $\bar{4}2m$ has the following form:

$$g_{ij} = \begin{vmatrix} g_{11} & 0 & 0 \\ 0 & -g_{11} & 0 \\ 0 & 0 & 0 \end{vmatrix}. \qquad (12)$$

Then the scalar gyration parameter may be represented as

$$G = g_{11} \sin^2 \Theta \cos^2 \varphi - g_{11} \sin^2 \Theta \sin^2 \varphi = g_{11} \sin^2 \Theta (\cos^2 \varphi - \sin^2 \varphi). \qquad (13)$$

The conoscopic patterns typical for propagation of light along z axis are depicted in Fig. 10. As seen from Fig. 10, the patterns for the case of $\Delta n_l \prec \Delta n_c$ reveal quite unusual shape: they are invariant with respect to the symmetry operation of four-fold axis. Moreover, the interference curves become closed and aperiodic if $\Delta n_l \succ 10^{-5}$. One can conclude that the region of rarefication of the interference curves corresponds to appearance of the mentioned special directions, which lie on a rather complicated surface, the one invariant with respect to the four-fold symmetry axis. With further increasing linear birefringence, the surface convolves into a single line oriented along z axis, which is in fact the ordinary optic axis in uniaxial crystals.

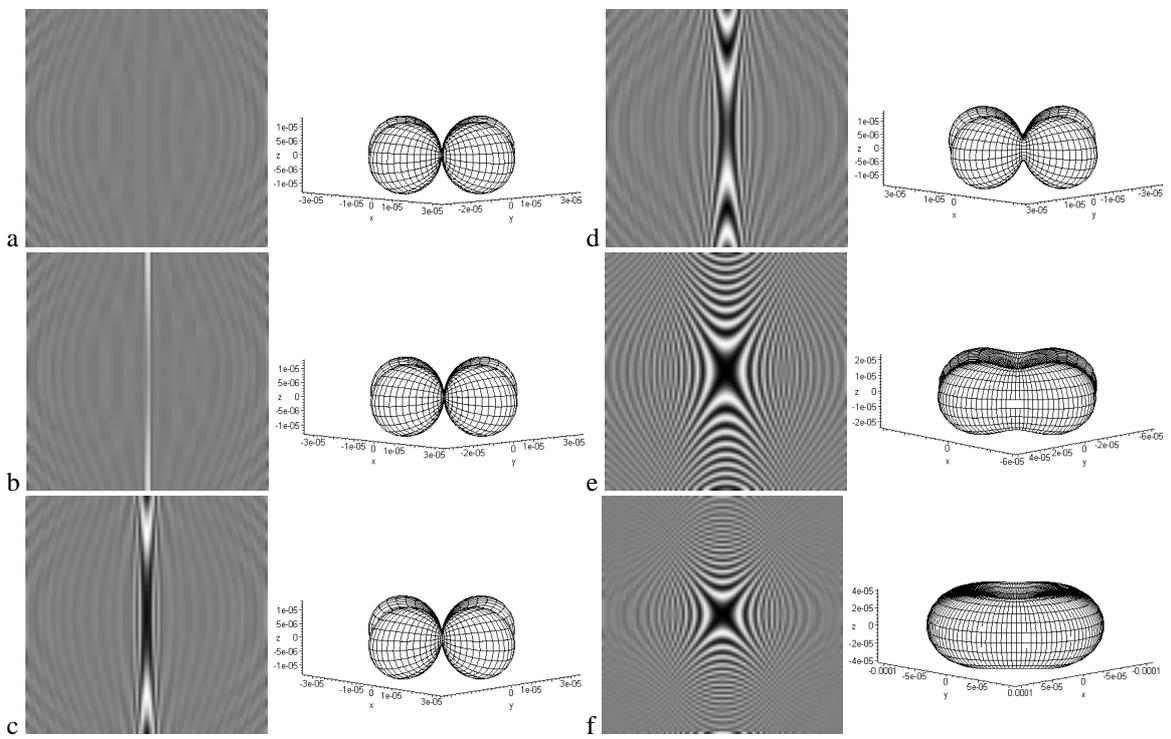

**Fig. 12.** Conoscopic patterns for the propagation direction characterized with the condition $G = \Delta n_c = 0$ ($45°$ with respect to x axis in the xy plane) in gyrotropic crystals of point group $\bar{4}2m$ at different values of linear birefringence and the corresponding surfaces of total elliptical birefringence: (a) $\Delta n_l = 0$, (b) $\Delta n_l = 10^{-6}$, (c) $\Delta n_l = 5 \times 10^{-6}$, (d) $\Delta n_l = 10^{-5}$, (e) $\Delta n_l = 5 \times 10^{-5}$ and (f) $\Delta n_l = 10^{-4}$ ($g_{11} = 5 \times 10^{-5}$, $\lambda = 632.8 nm$ and $d = 1m$).

*e) Propagation of light along x or y axes in the uniaxial crystals of point symmetry group $\bar{4}2m$.*

The conoscopic fringes peculiar for the propagation of light along x and y directions in crystals belonging to the point group of symmetry $\bar{4}2m$ are the same. When the light propagates along x or y directions under the conditions of $\Delta n_l = 0$, the interference fringes are ellipses. Those ellipses transform to hyperboles with increasing linear birefringence, through the intermediate circular and hyperbolic-elliptical states (see Fig. 11). Notice that the elliptical or circular shapes of the interference patterns occurring in this case mean that the surface of the total birefringence possesses the same symmetry as the conoscopic fringes.

*f) Propagation of light in the direction where $G = \Delta n_c = 0$ ($45°$ with respect to x axis in xy plane) in the uniaxial crystals of point symmetry group $\bar{4}2m$.*

When the light propagates directly along the so-called optic axis for circularly polarized waves (i.e., along the direction where $\Delta n_c = 0$), one can observe a transformation of interference fringes from an elliptic figure to the other one, through a straight interference band oriented parallel to z axis (Fig. 12a). Widening of this band and simultaneous movement of the special directions towards the z direction occur with increasing linear birefringence (see Fig. 12b–e). This conclusion is in agreement with the results shown in Fig. 10b–f. Once the linear birefringence becomes larger than the circular one, the usual conoscopic figure is observed (Fig. 12f).

## Discussion

Let us analyze some peculiarities of the special propagation directions revealed by us. When investigating the function

$$\Delta n = \sqrt{\Delta n_m^2 \sin^2 \Theta + \frac{g_{ij}^2 l_i^2 l_j^2}{n^2}} \tag{14}$$

for the existence of extrema (namely, the minima depending on the angles $\Theta$ and $\varphi$), one can easily find that the extreme values correspond just to the revealed special directions. Moreover, it is seen from the figures of total birefringence that the mentioned directions are also characterized by intersection of the surfaces of circular and linear birefringences and the equality of these birefringences:

$$\Delta n_m \sin \Theta = \pm \frac{g_{ij} l_i l_j}{\bar{n}}. \tag{15}$$

Then the ellipticity of the eigenwaves propagated along those directions under the condition of $\Delta n_c = \Delta n_l$ will acquire a specific value,

$$|k| = \frac{2\left(G/\bar{n}^5\right)}{B_2 - B_1 - \sqrt{(B_1 - B_2)^2 + 4\frac{G^2}{\bar{n}^8}}} = 0.414, \tag{16}$$

where $B_1$ and $B_2$ are the components of optical impermeability tensor actual for the cross section of the optical indicatrix perpendicular to the direction of light propagation. It is interesting to recall in this relation that the given eigenwave ellipticity has been revealed in the study [13], but, unfortunately, the authors [12,13] have not noticed that the value $|k| = 0.414$ should be peculiar for the condition $\Delta n_c = \Delta n_l$. Notice also that the value of the above eigenwave ellipticity can be achieved not only in the vicinity of the isotropic point, but also for the propagation of light along any direction where the condition $\Delta n_c = \Delta n_l$ is satisfied, i.e., somewhere far from the wavelength of the isotropic point. For example, the circular and linear birefringences in quartz crystals are the same if the light propagates in the direction defined by the angle $5°7'$ with respect to the optic axis [6,13,14] and we have $|k| = 0.414$, too.

**Conclusion**

The conoscopic patterns calculated for gyrotropic crystals in the vicinity of their isotropic point display a number of distinctive features, which can be summarized as the appearance of special directions, where the total elliptical birefringence acquires extreme values and where the linear and circular birefringences are equal to each other. Those directions lie on the surface of cone, which can convolve into one axis parallel to z direction in optically uniaxial crystals with increasing linear birefringence or, alternatively, the angle of this cone can then approach some minimal value determined by the ratio of linear and circular birefringences. In the case of crystals belonging to the point symmetry group $\bar{4}2m$, the surface of those directions is rather complicated and it reveals such the symmetry element as the four-fold axis, which coincides with z direction. Summing up the results of the present study, we have found a new peculiarity of the propagation of light in gyrotropic crystals characterized with commensurable values of the linear and circular birefringences: the appearance of special directions, for which the circular and linear birefringences become the same. Moreover, the above directions are characterized by a definite value of the eigenwave ellipticity, $|k| = 0.414$.

The results of further studies of this phenomena will be described in a forthcoming paper.

**Acknowledgement**

The authors acknowledge financial support of this study from the Ministry of Education and Science of Ukraine (the Project N0106U000615)